\documentclass{aa}
\usepackage{graphicx}
\usepackage{txfonts}
\usepackage{multirow}
\usepackage{color}
\usepackage{url}
\usepackage{amsmath}
\usepackage{amssymb}

\usepackage{natbib}
\bibpunct{(}{)}{;}{a}{}{,}

\def\be{\begin{equation}}
\def\ee{\end{equation}}
\def\bea{\begin{eqnarray}}
\def\eea{\end{eqnarray}}
\def\ba#1\ea{\begin{align}#1\end{align}}

\def\mr{\mathrm}
\def\lbra{\left\langle}
\def\rbra{\right\rangle}
\def\Cov{\mr{Cov}}
\def\Msun{M_{\odot}}

\def\smallminus{\,\text{-}\,}

\begin{document}

\title{Inadequacy of internal covariance estimation for super-sample covariance}
\titlerunning{Inadequacy of internal covariances for SSC}

%\author{Fabien Lacasa\inst{\ref{inst1}}\thanks{fabien.lacasa@unige.ch}
%\and  Martin Kunz\inst{\ref{inst1}}

\author{Fabien Lacasa\thanks{fabien.lacasa@unige.ch}
\and  Martin Kunz
}
\institute{
D\'{e}partement de Physique Th\'{e}orique and Center for Astroparticle Physics, Universit\'{e} de Gen\`{e}ve, 24 quai Ernest Ansermet, CH-1211 Geneva, Switzerland\label{inst1}
}

\date{\today}

\abstract
{
We give an analytical interpretation of how subsample-based internal covariance estimators lead to biased estimates of the covariance, due to underestimating the super-sample covariance (SSC). This includes the jackknife and bootstrap methods  as estimators for the full survey area, and subsampling as an estimator of the covariance of subsamples. The limitations of the jackknife covariance have been previously presented in the literature because it is effectively a rescaling of the covariance of the subsample area. However we point out that subsampling is also biased, but for a different reason: the subsamples are not independent, and the corresponding lack of power results in SSC underprediction. We develop the formalism in the case of cluster counts that allows  the bias of each covariance estimator to be exactly predicted. We find significant effects for a  small-scale area or when a low number of subsamples is used, with auto-redshift biases ranging from 0.4\% to 15\% for subsampling and from 5\% to 75\% for jackknife covariance estimates. The cross-redshift covariance is even more affected;  biases range from 8\% to 25\% for subsampling and from 50\% to 90\% for jackknife. Owing to the redshift evolution of the probe, the covariances cannot be debiased by a simple rescaling factor, and an exact debiasing has the same requirements as the full SSC prediction. These results thus disfavour the use of internal covariance estimators on data itself or a single simulation, leaving analytical prediction and simulations suites as possible SSC predictors.
}
\keywords{large-scale structure of the universe - methods: analytical}

\maketitle

%%%%%%%%%%%%%%%%%%%%%%%%%%%%%%%%%%%%%%%%%%%%%%%%%%%%%%%%%%%%%%%%%%%

\section{Introduction}\label{Sect:intro}

Cosmology is currently in a golden era of surveys probing an ever larger  fraction of the Hubble volume. In the next decade we will obtain data covering most of the extragalactic sky out to a redshift of two or more, thanks to surveys like the Dark Energy Survey \citep[DES,][]{Abbott:2005bi}, the Dark Energy Spectroscopic Instrument \citep[DESI,][]{Aghamousa:2016zmz}, Euclid \citep{Laureijs:2011gra}, the Large Synoptic Sky Telescope \citep[LSST,][]{Abell:2009aa}, and the Square Kilometer Array \citep[SKA,][]{Maartens:2015mra}, to name just a few. These surveys will try to answer some of the most challenging questions in cosmology and fundamental physics, about the nature of  dark matter and dark energy and the origin of the large-scale structure of the Universe \citep[e.g.\ ][and the survey references given above]{Amendola:2012ys}. Thanks to the sheer number of objects that will be observed in these surveys, the statistical uncertainties will be strongly reduced. However, in order to obtain accurate statistical inferences,  the systematic errors also need to be under control. A key quantity for the likelihood function is the survey covariance matrix of the observables. 
Incorrect covariance matrices, for example including only Gaussian contributions, can significantly overestimate the information content of galaxy surveys observables \citep{Repp2015}.

As galaxy surveys become deeper and larger, traditional sources of statistical errors such as shot noise are beaten down. The effect of super-survey modes, called super-sample covariance (SSC), however, decreases more slowly with the survey volume. These modes change the average matter density within the survey volume to which the probes react through linear (for 1 point probes such as cluster counts) or non-linear (for 2+ point probes such as the galaxy power spectrum) perturbation theory. The SSC effect was first pointed out for cluster counts by \cite{Hu2003}, and then for the matter power spectrum by \cite{Hamilton2006}. Super-sample covariance is particularly important when combining probes as it couples the observables together \citep{TakadaBridle2007, Takada2014, Krause2016, Lacasa2016}. Careful estimation of covariances including this term is thus necessary for single probe and combined probes analyses with current and future surveys.

Survey covariances can be estimated from the data itself through jackknife or bootstrap methods. However \cite{Hoffmann2015} showed that the standard jackknife estimator failed to recover correctly the 3D cluster counts of the MICE simulations and proposed an improved estimator instead. For galaxy weak lensing, \cite{Friedrich2016} showed that bootstrap and jackknife recover only partially the correct volume of cosmological parameter constraints, and \cite{Shirasaki2016} further pointed out the limitations of the jackknife covariance in the regime dominated by super-sample covariance (SSC).

In their standard form these methods  fail because they intrinsically assume that the subsamples are independent draws of a given random field. However, in reality the subsamples are not independent because they are correlated by large-scale fluctuations, which is exactly the SSC effect.

This article gives the formalism to exactly predict how biased the jackknife and bootstrap covariance estimators are, further driving home the point that they should not be used for cosmological parameter inference. It also examines the case of the subsampling method, which was designed to estimate the covariance of the subsample area instead of the full survey area. This case arises for example when cutting down a large simulation into survey-shaped pieces, and we show that this also produces a biased covariance. The capacity of predicting analytically this bias is of importance when  validating a prediction code against a single numerical simulation or a limited set of them. Indeed, we do not now possess  the resources to produce a large number of full observable universe sized simulations with all the necessary cluster, galaxy, and weak-lensing physics for current and future surveys, though smaller simulations with adequate rescaling have  recently been proposed as an alternative \citep{Klypin2017}.

This article is organised as follows: we review the internal covariance estimators in Sect. \ref{Sect:intern-cov-estim}; we derive analytically the impact of the SSC on these estimators in Sect. \ref{Sect:effect-on-SSC}; we  apply this formalism numerically to quantify the importance of the effect in Sect. \ref{Sect:num-results};  and finally we discuss our results and their perspectives in Sect. \ref{Sect:discussion}.

%Extra facts
In all numerical computations, we take a flat $\Lambda$CDM cosmology with parameters  
$h=0.67$, $\Omega_b h^2=0.022$, $\Omega_c h^2=0.12$, $n_S=0.96$ , $\sigma_8=0.83$.
The cluster counts are computed in two bins of redshift in the range $z\in[0.9,1.1]$ with a width $\Delta z=0.1$, and four bins of mass in the range $\log [M/(h^{-1}\Msun)] \in[14,16]$ 
with a width $\Delta\log [M/(h^{-1}\Msun)]=0.5$. The corresponding covariance is thus an $8\times8$ matrix composed of $4\times4$ auto- and cross-redshift blocks. The halo mass function is taken from \cite{Tinker2008}, with the corresponding halo bias from \cite{Tinker2010}, and the linear matter power spectrum is from the transfer function by \cite{Eisenstein1998}. We quote these details here for completeness, but the precise values do not affect the conclusions of this paper in a qualitative way.

%%%%%%%%%%%%%%%%%%%%%%%%%%%%%%%%%%%%%%%%%%%%%%%%%%%%%%%%%%%%%%%%%%%%
\section{Internal covariance estimators}\label{Sect:intern-cov-estim}

Let us denote by an index $\alpha$ the bins of cluster mass and redshift: $\alpha\sim(i_M,i_z)$. The survey area is divided in $N_\mr{sub}$ equal-geometry subsamples with a latin index $i$. As will be shown in the subsections below, any natural covariance estimators built from these subsamples can be written in the form
\be\label{Eq:Cov-wij}
\widehat{\Cov}(N_\alpha,N_\beta) = \sum_{i,j=1}^{N_\mr{sub}} w_{i,j} \ N_\alpha(i) \ N_\beta(j)
\ee
with the requirement that
\be
\sum_{i,j=1}^{N_\mr{sub}} w_{i,j} = 0
,\ee
which implements the constraint that $\widehat{\Cov}=0$ if $N(i)=\mr{cst}$, i.e. the covariance estimator should yield zero if all samples are identical.

The main point is that all covariance estimators below are purely `survey internal' and are unable to see modes on scales larger than the survey volume. Because of this limitation they are unable to correctly take into account the super-sample covariance.

%%%%%%%%%%
\subsection{Subsampling}\label{Sect:subsamp}
The subsampling method tries to estimate the covariance of the subsample area through the simple empirical covariance of the subsample counts:
\be
\widehat{\Cov}_{\rm sub}(N_\alpha,N_\beta) = \frac{1}{N_\mr{sub}\smallminus 1} \sum_{i=1}^{N_\mr{sub}} \left(N_\alpha(i) \smallminus \overline{N}_\alpha\right) \left(N_\beta(i) \smallminus \overline{N}_\beta\right)
\ee
with
\be
\overline{N}_\alpha \equiv \frac{1}{N_\mr{sub}} \sum_{i=1}^{N_\mr{sub}} N_\alpha(i) \, .
\ee
This leads to an estimator of the form of Eq.\ (\ref{Eq:Cov-wij}) with
\be
w_{i,j}^\mr{sub} = \frac{\delta_{i,j} - \frac{1}{N_\mr{sub}}}{N_\mr{sub}-1} \, .
\ee

%%%%%%%%%%
\subsection{Jackknife}\label{Sect:jackknife}
The jackknife method, however, aims to recover the covariance of the full survey area. To do so, it defines a jackknife sample as the total survey minus one subsample. Denoting $N_\alpha^{\rm jack}(i) = N_\alpha^{tot} - N_\alpha(i)$ such jackknife sample, we have
\begin{eqnarray}
\lefteqn{\widehat{\Cov}_{\rm jack}(N_\alpha,N_\beta) =} && \nonumber \\
&&\frac{N_\mr{sub}}{N_\mr{sub}\smallminus 1} \sum_{i=1}^{N_\mr{sub}} \left(N^{jk}_\alpha(i) \smallminus \overline{N}^{\rm jack}_\alpha\right) \left(N^{jk}_\beta(i) \smallminus \overline{N}^{\rm jack}_\beta\right)
,\end{eqnarray}
where
\ba
\overline{N}^{\rm jack}_\alpha &\equiv \frac{1}{N_\mr{sub}} \sum_{i=1}^{N_\mr{sub}} N^{jk}_\alpha(i)\\
&= N_\alpha^{\rm tot} - \overline{N}_\alpha \, .
\ea
It easily follows that
\ba
\widehat{\Cov}_{\rm jack}(N_\alpha,N_\beta) &= \frac{N_\mr{sub}}{N_\mr{sub} \smallminus 1} \sum_{i=1}^{N_\mr{sub}} \left(N_\alpha(i) \smallminus \overline{N}_\alpha\right) \left(N_\beta(i) \smallminus \overline{N}_\beta\right)\\
&= N_\mr{sub} \times \widehat{\Cov}_{\rm sub}(N_\alpha,N_\beta)
\ea
and thus the jackknife estimator is just a rescaling of the subsampling estimator. It can thus be put in the form Eq.~(\ref{Eq:Cov-wij}) with
\be
w_{i,j}^\mr{jack} = N_\mr{sub} \ \frac{\delta_{i,j} - \frac{1}{N_\mr{sub}}}{N_\mr{sub}-1} \, .
\ee

%%%%%%%%%%
\subsection{Bootstrap}\label{Sect:bootstrap}
Like the jackknife method described in Sect. \ref{Sect:jackknife}, the bootstrap method attempts to estimate the covariance of the full survey area. To do so, it first defines a bootstrap sample $b$ as the union of $N_\mr{sub}$ possibly repeated subsamples. For example with $N_\mr{sub}=4$, a possible sample is \{3,1,4,1\}. A high number of samples $N_\mr{bs}$ is drawn, typically 1,000 or 10,000, and the natural empirical covariance estimator is applied on these samples:
\ba
\lefteqn{\widehat{\Cov}_{\rm boot}(N_\alpha,N_\beta) =} && \nonumber \\
&&\frac{1}{N_\mr{boot} \smallminus 1} \sum_{b=1}^{N_\mr{boot}} \left(N^{bs}_\alpha(b) \smallminus \overline{N}^{\rm boot}_\alpha\right) \left(N^{\rm boot}_\beta(b) \smallminus \overline{N}^{bs}_\beta\right) \, .
\ea
In Appendix \ref{App:bootstrap}, it is shown that when $N_\mr{boot}\rightarrow\infty$, the bootstrap method basically asymptotes to the jackknife one, giving an estimator of the form Eq.~(\ref{Eq:Cov-wij}) with
\be
w_{i,j}^\mr{boot} = \delta_{i,j} - \frac{1}{N_\mr{sub}}
,\ee
which is the same as the jackknife weight, up to a factor $\frac{N_\mr{sub}}{N_\mr{sub}-1}\simeq 1$.\\
We note that this factor can be cancelled by applying an alternative normalisation to the jackknife estimator, and it has an impact smaller than the effects discussed in the following sections of this article. Thus, we will not be mentioning the bootstrap method again, just the subsampling and jackknife methods.\\

It should be emphasized that the estimators discussed here all work correctly in their respective domains of applicability. The main issue is that in the presence of SSC the different subsamples are not independent and identically distributed, which breaks an important assumption that underlies these estimators.

%%%%%%%%%%%%%%%%%%%%%%%%%%%%%%%%%%%%%%%%%%%%%%%%%%%%%%%%%%%%%%%%%%%%
\section{Effect on super-sample covariance}\label{Sect:effect-on-SSC}

\begin{figure}[!th]
\begin{center}
\includegraphics[width=.49\linewidth]{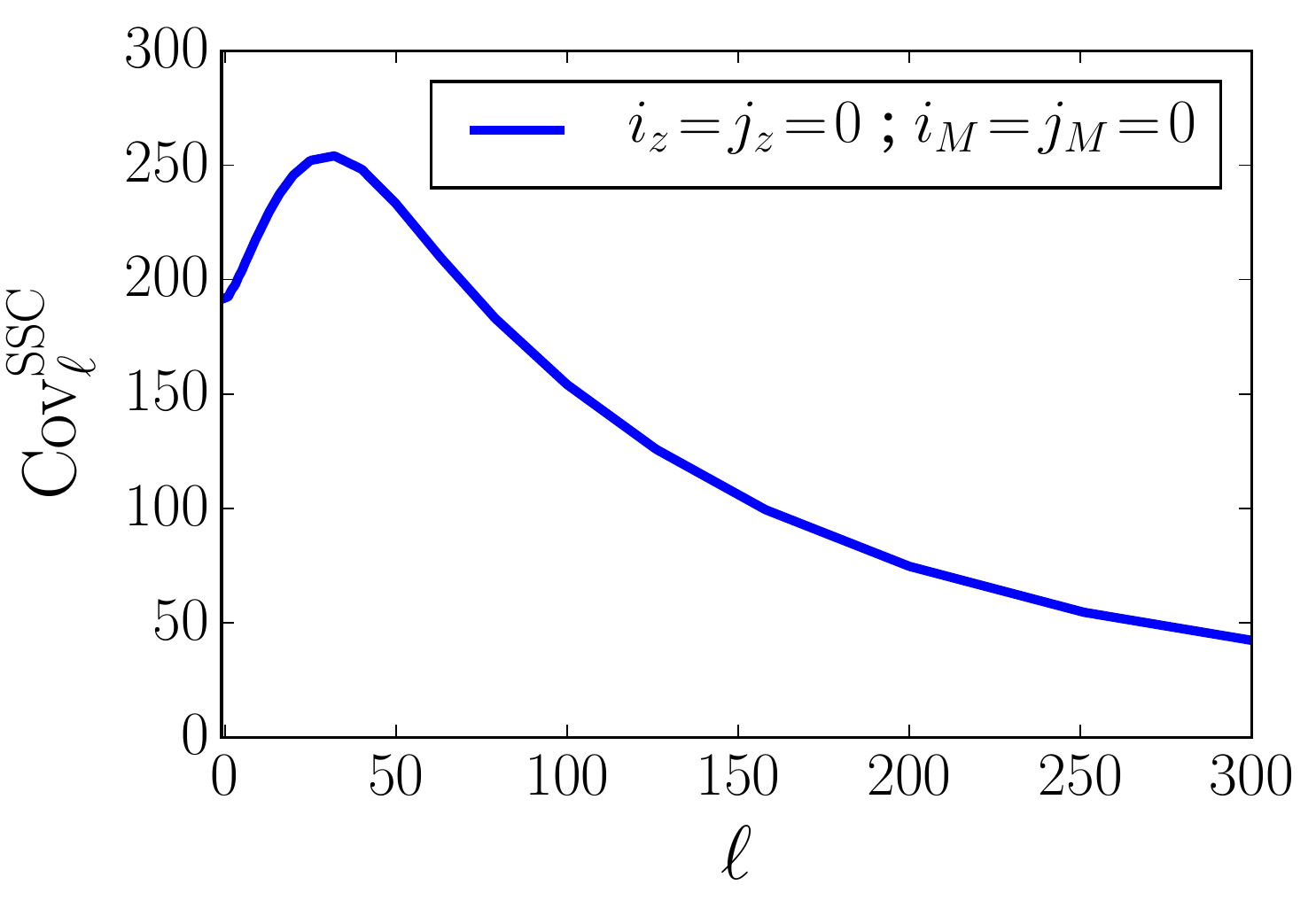}
\includegraphics[width=.49\linewidth]{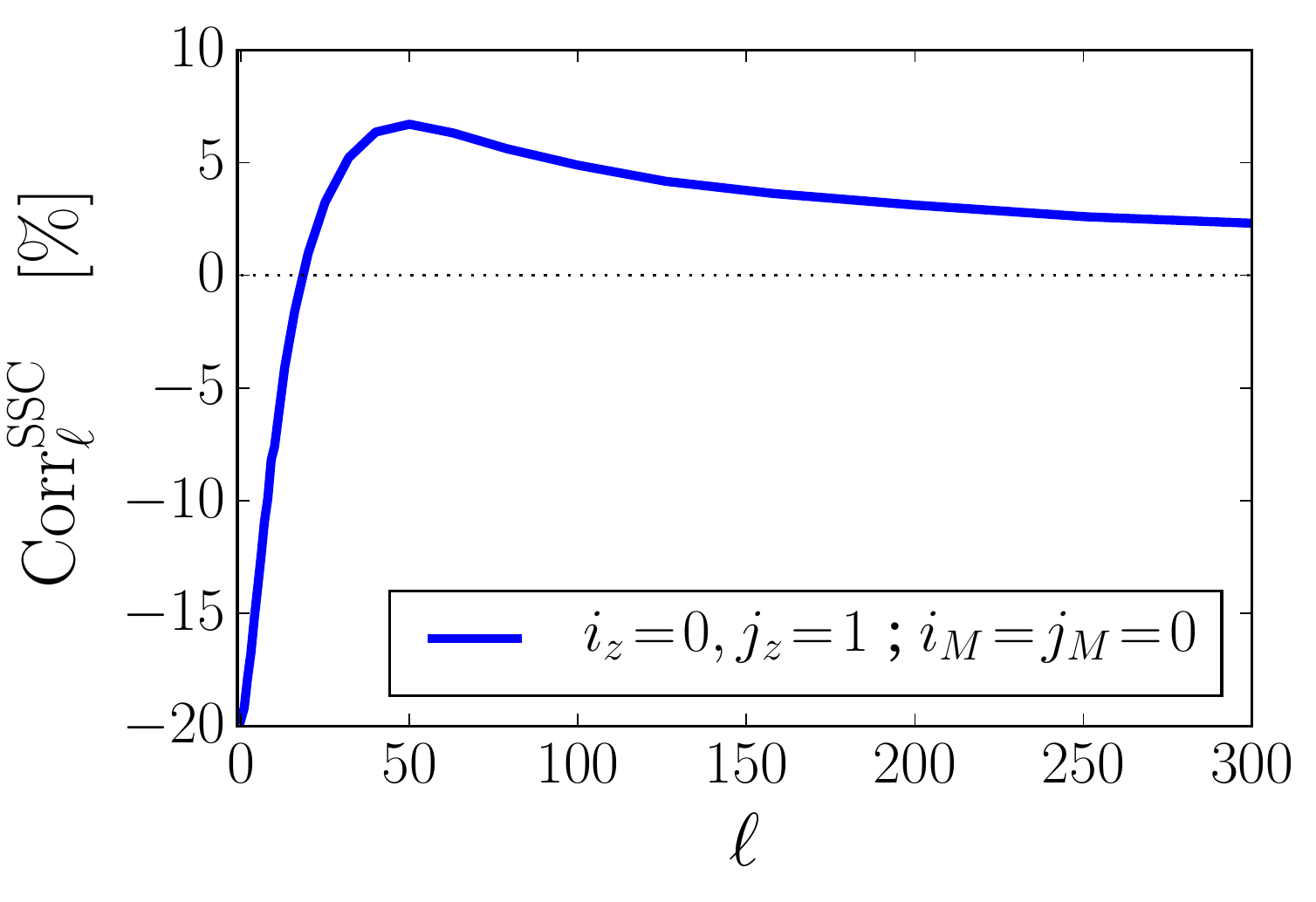}
\caption{Multipole dependence of $\Cov_\ell^\mr{SSC}$. \textit{Left:} Auto-redshift covariance. \textit{Right:} Cross-redshift correlation.}
\label{Fig:Cov_ell^SSC}
\end{center}
\end{figure}

Following \cite{Lacasa2016b}, the super-sample covariance of the number counts in a given subsample $i$ can be predicted as
\ba\label{Eq:Cov-counts-wcovell}
\nonumber \lbra N_\alpha(i) \, N_\beta(i) \rbra_c &\equiv \lbra N_\alpha(i) \, N_\beta(i) \rbra - \lbra N_\alpha(i)\rbra \lbra N_\beta(i) \rbra\\
\nonumber &= \frac{1}{f^2_\mr{sky}(i)} \sum_\ell \frac{2\ell+1}{4\pi} \ C_\ell\left(M(i)\right) \\
& \qquad \times \Cov_\ell^\mr{SSC}\left(N_\alpha , N_\beta\right),
\ea
where $M(i)$ is the mask corresponding to the $i$th subsample and $f_\mr{sky}(i)$ is the corresponding sky fraction. The factor $\Cov_\ell^\mr{SSC}$, which is independent of geometry but does depend on the cosmology and general modelling, is plotted in Fig. \ref{Fig:Cov_ell^SSC}. The plot shows the $\ell$ dependence of this factor for an arbitrary and unimportant choice of mass bins, for the auto-redshift covariance and the cross-redshift correlation (i.e. covariance normalised by its diagonal).

For the auto-redshift covariance, $\Cov_\ell^\mr{SSC}$ exhibits a behaviour similar to the matter power spectrum which sources it, first increasing with multipole up to a maximum corresponding to the matter-radiation equality angular scale at that redshift ($\ell_\mr{eq} = k_\mr{eq}/r(z)$) and then decreasing monotonically. The cross-redshift covariance shows a large anti-correlation on large scales (reaching -20\% for a full-sky survey for which $C_\ell(M)\propto \delta_{\ell,0}$),  then changes  sign and  reaches a maximum of 7\% at $\ell=50$, and then decreases to zero reaching 2.5\% at $\ell=300$.

Formula~(\ref{Eq:Cov-counts-wcovell}) can straightforwardly be generalised to predict the cross-covariance between two different subsamples $i$ and $j$, involving the cross-spectrum of their respective masks $C_\ell\left(M(i),M(j)\right)$:
\ba
\nonumber \lbra N_\alpha(i) \, N_\beta(j) \rbra_c &= \frac{1}{f_\mr{sky}(i) \, f_\mr{sky}(j)} \sum_\ell \frac{2\ell+1}{4\pi} \ C_\ell\left(M(i),M(j)\right) \\
& \qquad \times \Cov_\ell^\mr{SSC}\left(N_\alpha , N_\beta\right) \, .
\ea
In the following, we  assume that all subsamples have the same area, denoted $f_\mr{sky}^\mr{sub}$.

It follows from the previous equations that the average value of a covariance estimator of the form of Eq.~(\ref{Eq:Cov-wij}) is
\ba\label{Eq:avgcov-cleff}
\nonumber \lbra \widehat{\Cov}(N_\alpha,N_\beta) \rbra &= \sum_{i,j=1}^{N_\mr{sub}} w_{i,j} \left( \lbra N_\alpha(i) \ N_\beta(j) \rbra_c + \overline{N}_\alpha \, \overline{N}_\beta\right)\\
&= \frac{1}{\left(f^\mr{sub}_\mr{sky}\right)^2} \sum_\ell \frac{2\ell+1}{4\pi} \ C_\ell^\mr{eff} \ \Cov_\ell^\mr{SSC}\left(N_\alpha , N_\beta\right),
\ea
where the effective power spectrum is
\be\label{Eq:def-cleff}
C_\ell^\mr{eff} = \sum_{i,j=1}^{N_\mr{sub}} w_{i,j} \ C_\ell\left(M(i),M(j)\right) \, .
\ee
The difference between the effective power spectrum and the real power spectrum of the survey area is the exact reason that makes internal covariance estimators biased for SSC. This can be regarded as the main point of this article.

%%%%%%%%%%%%%%%%%%%%%%%%%%%%%%%%%%%%%%%%%%%%%%%%%%%%%%%%%%%%%%%%%%%%
\section{Numerical results}\label{Sect:num-results}
In this section we evaluate  Eq. (\ref{Eq:avgcov-cleff}) numerically with the effective power spectrum given by Eq. (\ref{Eq:def-cleff}) and compare it to the true covariance using the true power spectrum in order to quantify the possible bias of the covariance estimator.

%%%%%%%%%%
\subsection{Setup}

We defined two subsampling setups for illustration in this article. In both cases, we chose the subsamples as Healpix low resolution pixels as it allows for fast Healpix operations and gives the subsamples  the exact same geometry and area.

\begin{figure}[!th]
\begin{center}
\includegraphics[width=.9\linewidth]{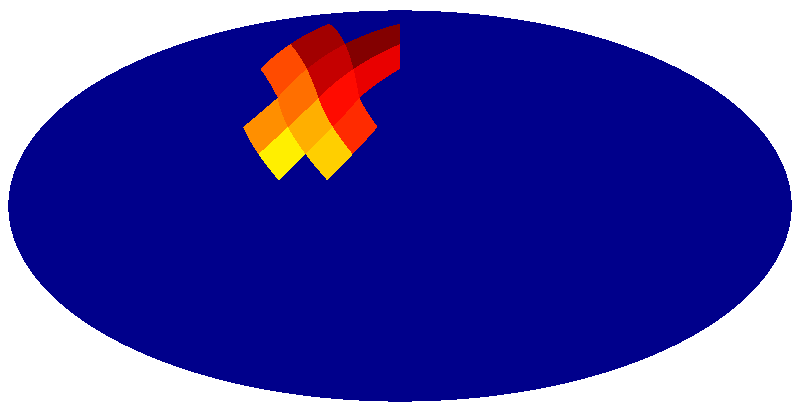}
\includegraphics[width=.6\linewidth]{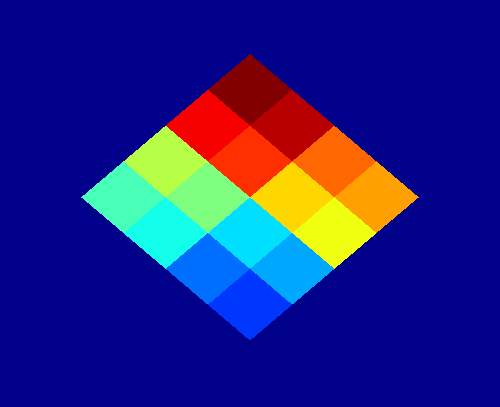}
\caption{Illustration of the subsamples used in the analysis. The colour code is arbitrary, and only serves to  differentiate between the subsamples. \textit{Top:} Large-scale case. \textit{Bottom:} Small-scale case.}
\label{Fig:subsamples-illustration}
\end{center}
\end{figure}

In the first case, hereafter called the large-scale setup, we selected the maximum number of Healpix pixels at $N_\mr{side}=4$ that fit strictly in the octant $0<\theta<\pi/2$ $0<\phi<\pi/2$, and further removed the northernmost pixel\footnote{This pixel is unfit for $C_\ell$ computations because the Healpix pixelisation has poor sampling of $\phi$ near the poles. Other pixels are removed because they do not reside entirely within the octant, giving the total area  the X shape visible in Fig.~\ref{Fig:subsamples-illustration}.}. This setup is visible in the top panel of Fig. \ref{Fig:subsamples-illustration} and gives subsamples of area $\sim 215$ deg$^2$. This situation arises for covariance estimations from a numerical simulation. Indeed, with an N-body simulation, light cone observables are usually computed on a sky octant, which corresponds to putting the observer at one corner of the simulation box \cite{Sehgal2010,Fosalba2015}.\\

In the second case, herafter called the small-scale setup, we  divided a Healpix pixel at $N_\mr{side}=32$ in 16 subsamples corresponding to Healpix subpixels at $N_\mr{side}=128$. This setup is visible in the bottom panel of Fig. \ref{Fig:subsamples-illustration} and gives subsamples of area $\sim 0.2$ deg$^2$. This situation is a toy model of a small-scale rectangular survey.

%%%%%%%%%%
\subsection{Subsampling}

\begin{figure}[!th]
\begin{center}
\includegraphics[width=0.49\linewidth]{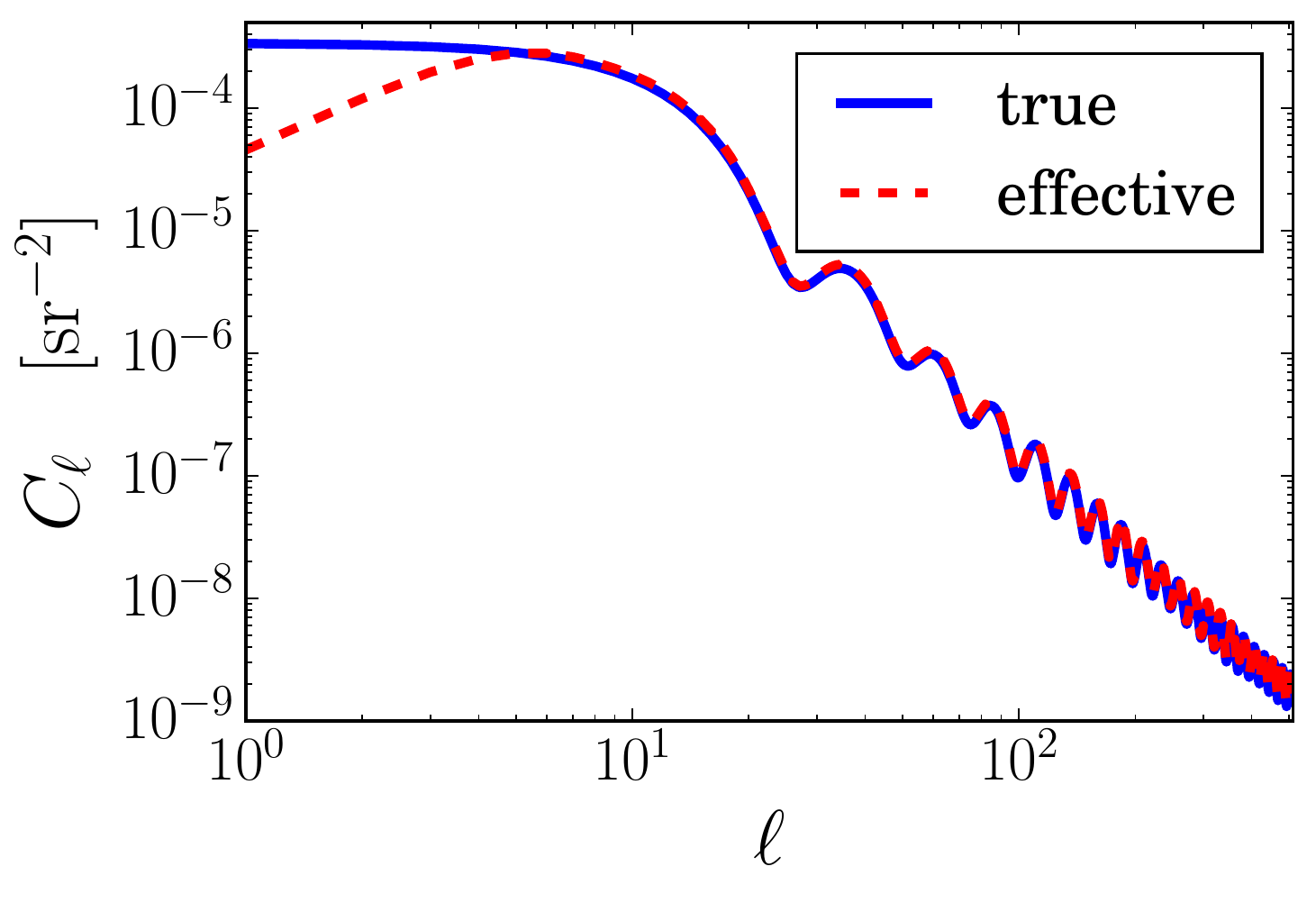}
\includegraphics[width=0.49\linewidth]{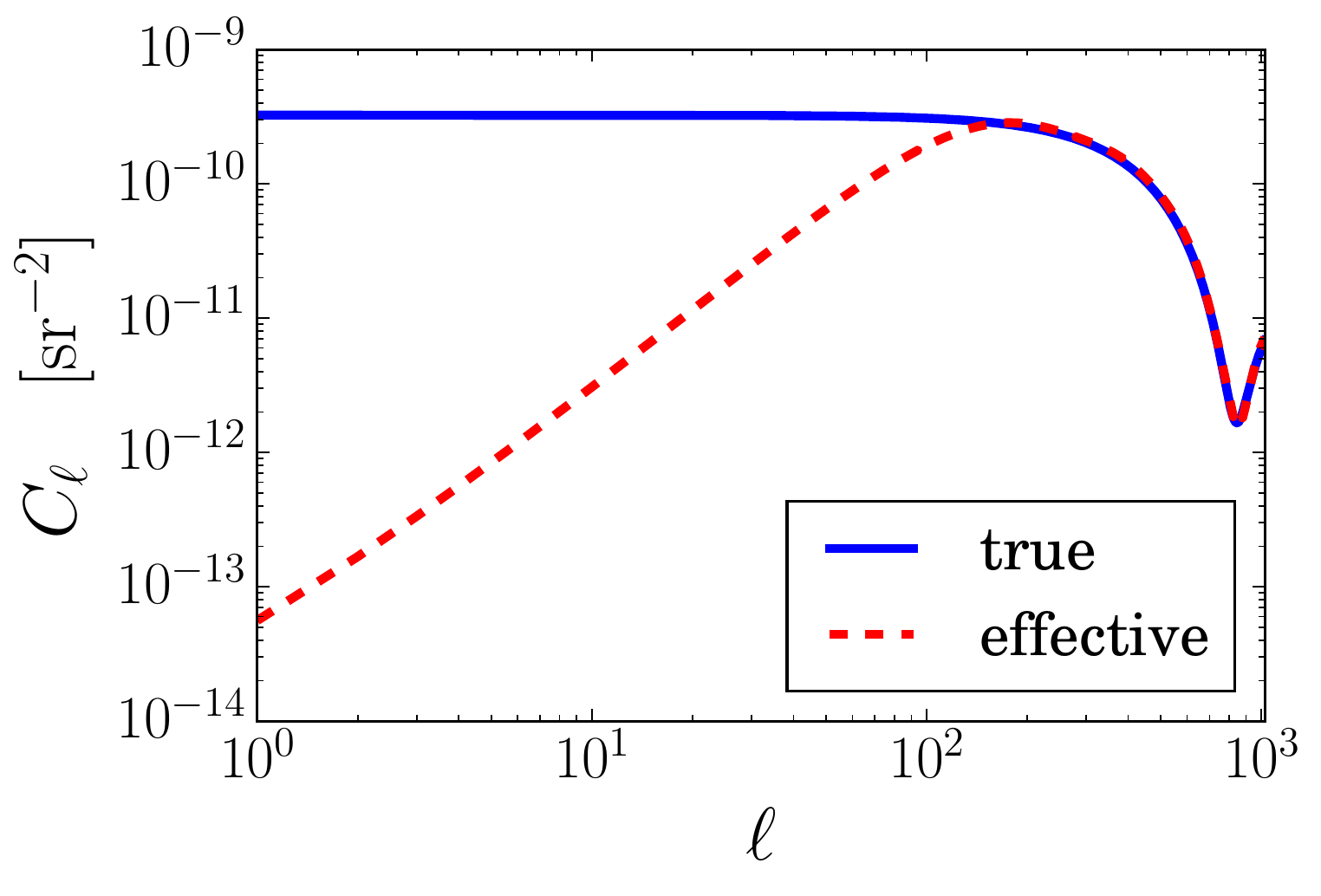}
\caption{True power spectrum of the subsample compared to the effective power spectrum that enters the subsampling covariance estimate. \textit{Left:} Large-scale case. \textit{Right:} Small-scale case.}
\label{Fig:cl-subvseff}
\end{center}
\end{figure}

Figure \ref{Fig:cl-subvseff} shows the effective power spectrum of the subsampling method, compared to the power spectrum of the subsample mask that is needed for the correct SSC prediction.

We see a lack of power in the effective power spectrum on scales comparable to or larger than the survey size. This corresponds to $\ell \lesssim 4$ in the large-scale case, and extends to $\ell \lesssim 150$ in the small-scale case.

\begin{figure}[!th]
\begin{center}
\includegraphics[width=0.49\linewidth]{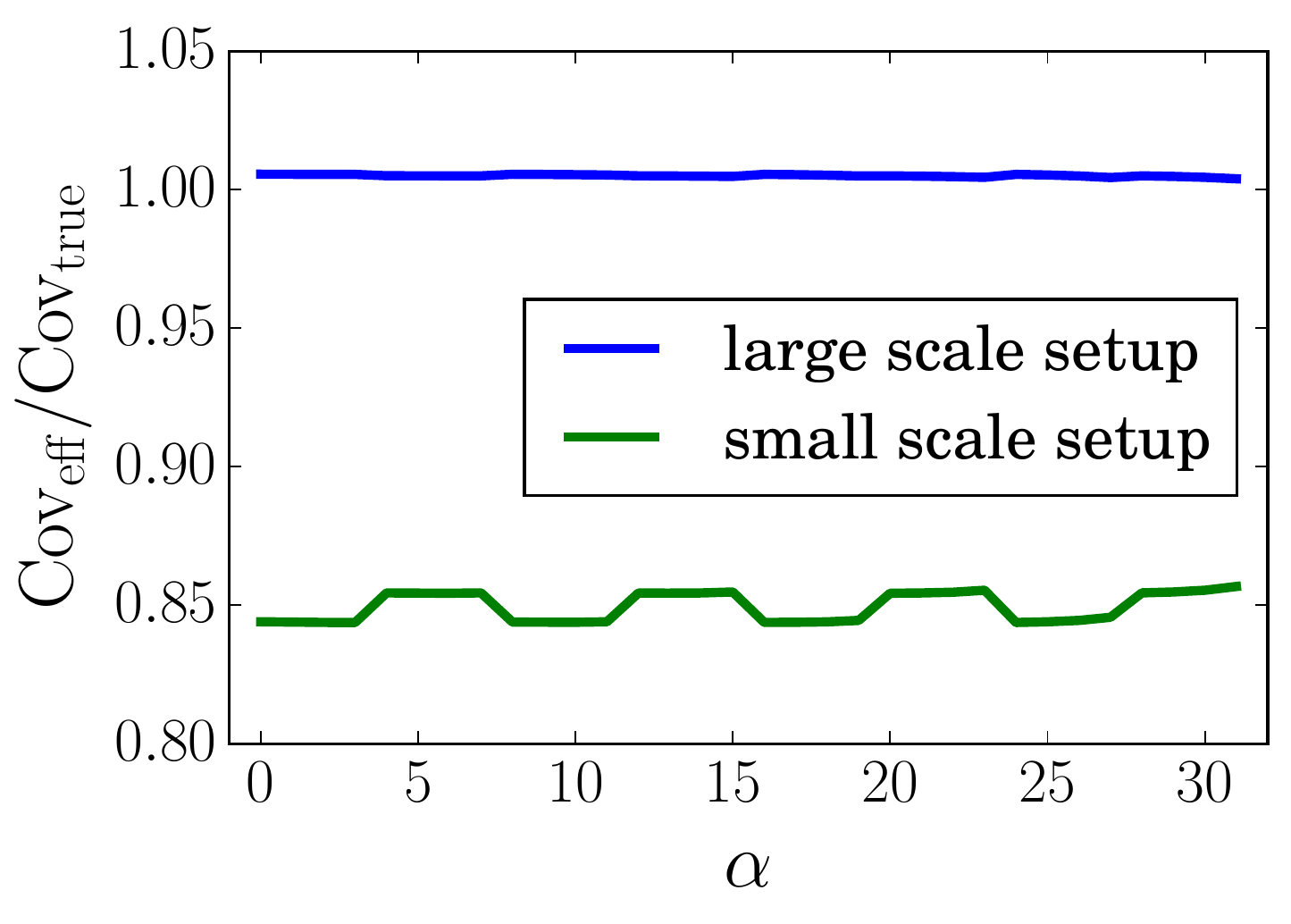}
\includegraphics[width=0.49\linewidth]{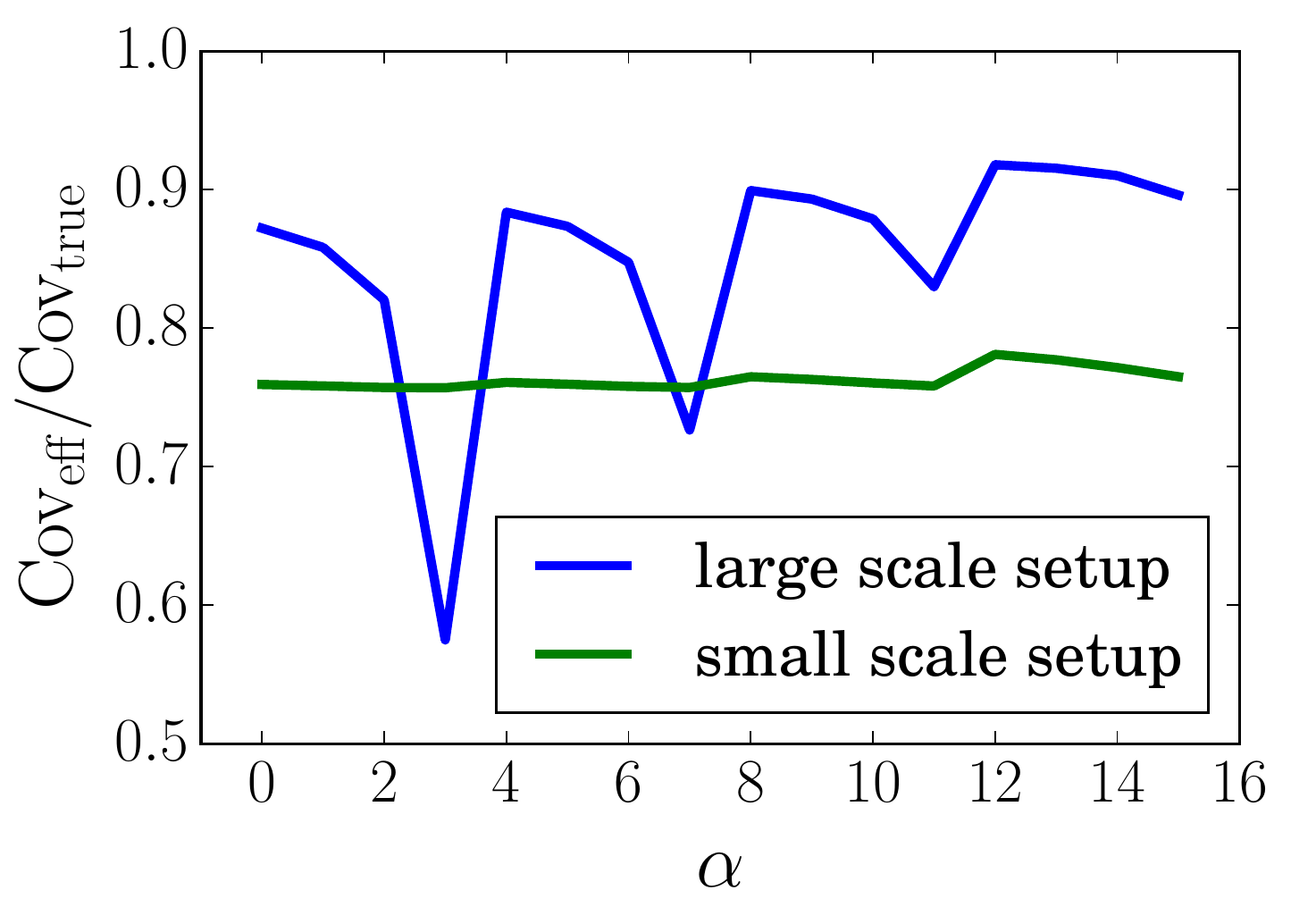}
\caption{Ratio of the effective covariance coming from subsampling to the true covariance of the subsample. \textit{Left:} Auto-redshift case. \textit{Right:} Cross-redshift case.}
\label{Fig:ratio-cov-eff/true-subsampling}
\end{center}
\end{figure}

We computed the corresponding cluster count covariances, and Fig.\ \ref{Fig:ratio-cov-eff/true-subsampling} shows the ratio of the subsampling covariance to the true one for both cases.

In the large-scale setup, the subsampling method correctly estimates the auto-redshift covariance. However the cross-redshift covariance is dominated by small multipoles,  visible in the right panel of Fig. \ref{Fig:Cov_ell^SSC}. The lack of power at small multipoles leads to an underestimate of the true cross-redshift covariance by 10\% to 40\% using the subsampling method. In this setup, the cross-$z$ covariance is negative and ranges between -1.5\% and -7.5\% when the covariance matrix is normalised to its diagonal.

In the small-scale setup, the lack of power on a large portion of the multipole range makes the subsampling method underestimate the covariance by $\sim15\%$ for auto-redshift and $\sim23\%$ for cross-redshift. In this setup, the cross-$z$ covariance is positive and ranges between 1.5\% and 5.5\% when the covariance matrix is normalised to its diagonal.

We note that in most cases, the difference between the estimated and true covariances is comparable to or larger than the 10\% systematic uncertainty due to the 5\% precision on halo bias predictions \citep{Hoffmann2017}.

%%%%%%%%%%
\subsection{Jackknife}

\begin{figure}[!th]
\begin{center}
\includegraphics[width=.49\linewidth]{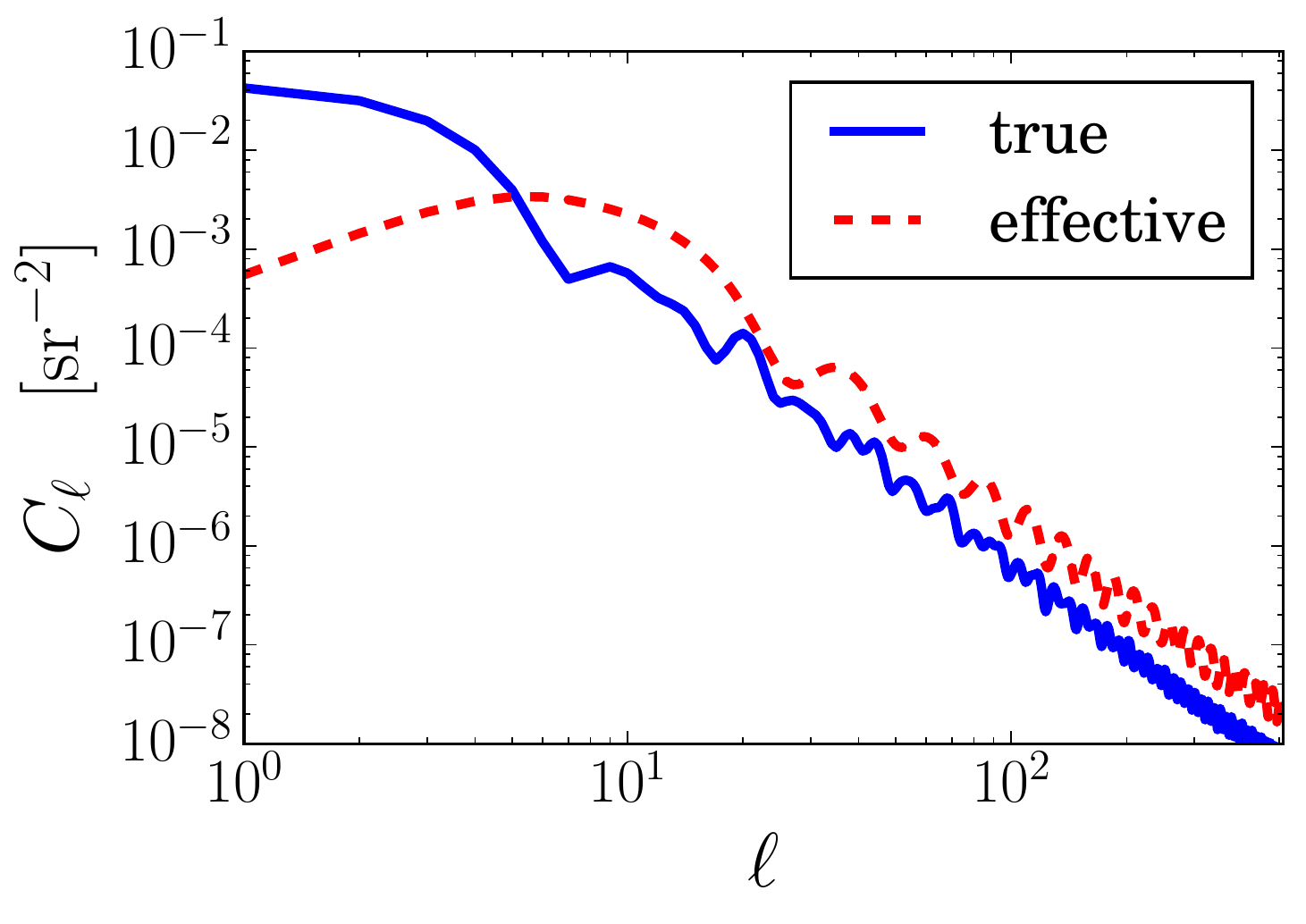}
\includegraphics[width=.49\linewidth]{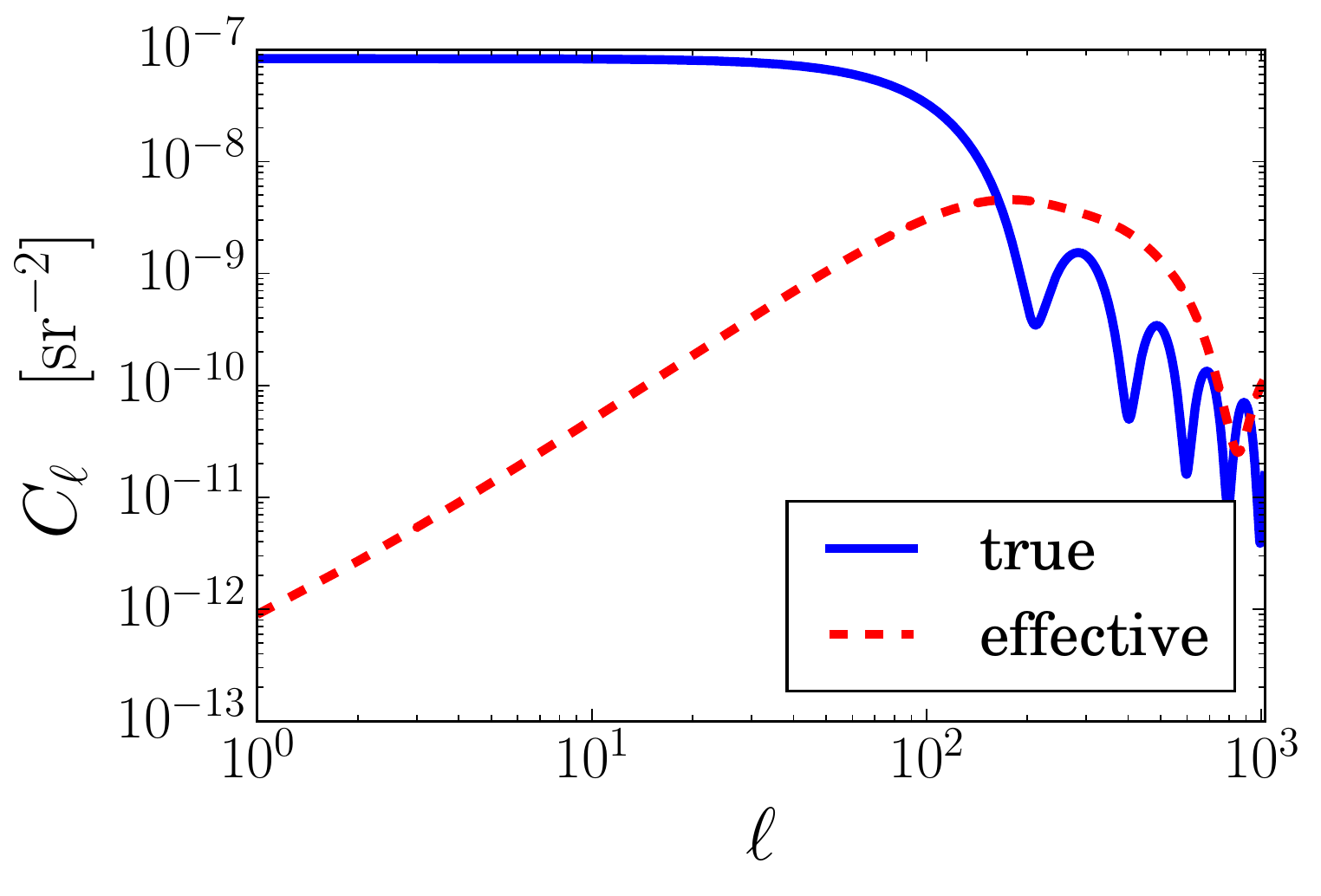}
\caption{True power spectrum of the survey area compared to the effective power spectrum that enters the jackknife covariance estimate. \textit{Left:} Large-scale case. \textit{Right:} Small-scale case.}
\label{Fig:cl-totvseff}
\end{center}
\end{figure}

Figure \ref{Fig:cl-totvseff} shows the effective power spectrum of the jackknife method, this time compared to the full survey area power spectrum.

In the large-scale setup, the jackknife power spectrum is larger than the true power spectrum of the survey on most of the multipole range, except for a lack of power in the few first multipoles. In the small-scale setup, however, the lack of power dominates most of the multipole range;   the jackknife power spectrum becomes larger than the true spectrum only on small scales.

\begin{figure}[!th]
\begin{center}
\includegraphics[width=0.49\linewidth]{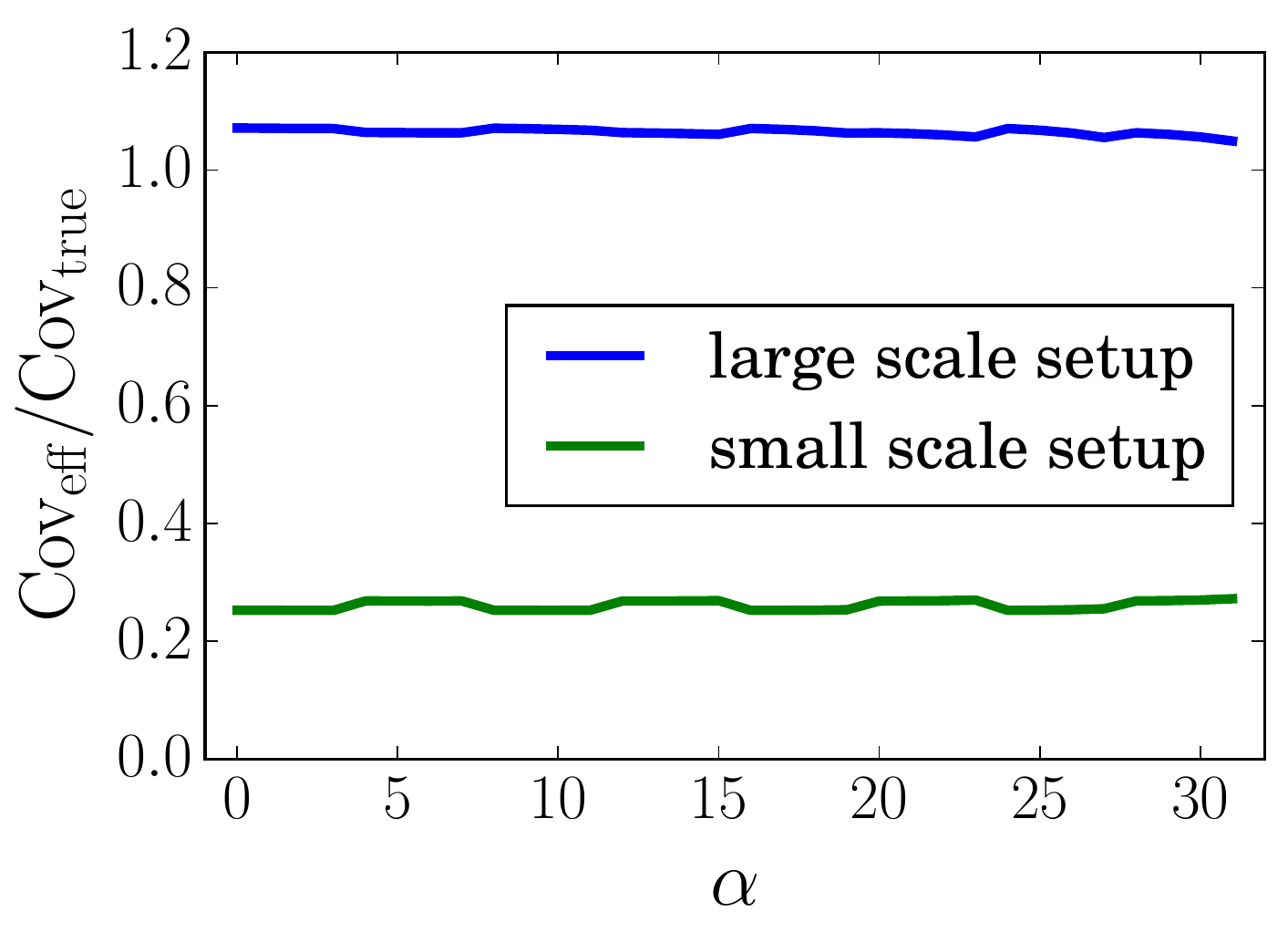}
\includegraphics[width=0.49\linewidth]{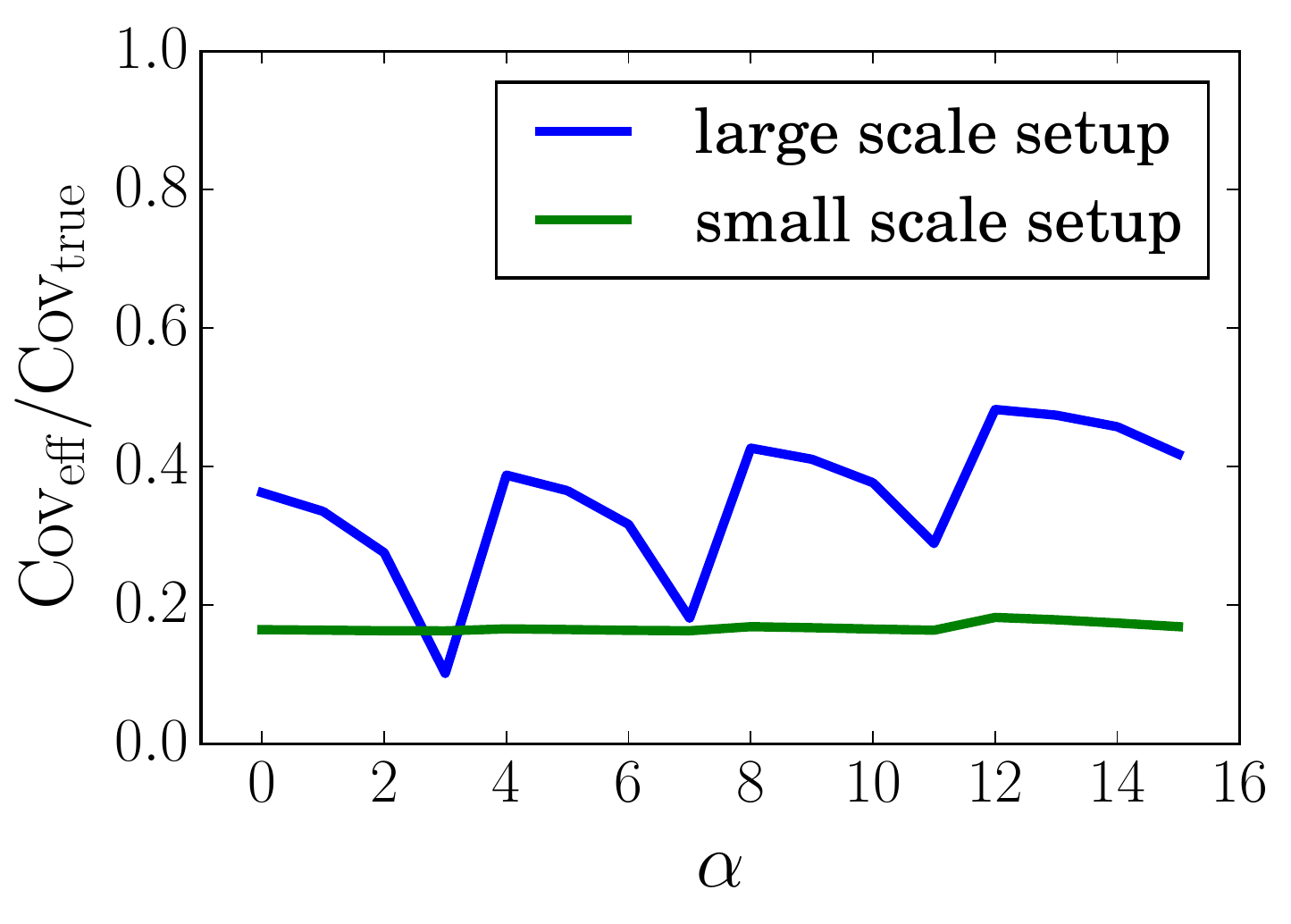}
\caption{Ratio of the effective covariance coming from jackknife to the true covariance of the total survey area. \textit{Left:} Auto-redshift case. \textit{Right:} Cross-redshift case.}
\label{Fig:ratio-cov-eff/true-jackknife}
\end{center}
\end{figure}

The corresponding cluster count covariances were again computed, and Fig.\ \ref{Fig:ratio-cov-eff/true-jackknife} shows the ratio of the jackknife covariance to the true one for both setups.

In the large-scale setup, the jackknife method overestimates the auto-redshift covariance by $\sim6\%$, due to the power spectrum excess. The cross-redshift covariance is dominated by small multipoles (see right panel of Fig. \ref{Fig:Cov_ell^SSC}), and we find that the jackknife method recovers less than half of the true covariance. In this setup, the cross-$z$ covariance is negative and ranges  between -9\% and -15\% when the covariance matrix is normalised to its diagonal. We note that cross-$z$ correlation can have a significant impact for constraints on cosmological parameters sensitive to the redshift evolution of the probe(s), such as the dark energy equation of state.

In the small-scale setup, the  lack of power on most of the multipole range makes the jackknife underestimate the covariance by a factor 3-4 for auto-redshift and 5-6 for cross-redshift. In this setup, the cross-$z$ covariance is positive and ranges  between 2\% and 8\% when the covariance matrix is normalised to its diagonal.

We note that in most cases, the difference between the estimated and true covariances is again comparable to or larger than the 10\% systematic uncertainty due to the 5\% precision on halo bias predictions \citep{Hoffmann2017}.

%%%%%%%%%%%%%%%%%%%%%%%%%%%%%%%%%%%%%%%%%%%%%%%%%%%%%%%%%%%%%%%%%%%%
\section{Discussion}\label{Sect:discussion}

\subsection{Results}

As has been shown previously in the literature \citep{Hoffmann2015,Shirasaki2016,Friedrich2016}, the standard jackknife method gives a biased estimate of the covariance of the survey area. This arises because the jackknife is in fact a constant rescaling of the natural estimator of the subsample area. This rescaling is calibrated to give an unbiased estimate of the Poissonian shot-noise; however, it cannot reproduce super-sample covariance because the latter has a non-trivial dependence on the mask geometry which depends on cosmological parameters \cite{Lacasa2016b}. 

We have given here an analytical understanding of this phenomenon and developed the formalism to predict the covariance bias quickly and without any simulation. We emphasize that the numerical calculations were performed on a simple dual-core laptop in at most tens of seconds. This article thus offers a fast way for survey analysts to check how biased their favourite internal covariance estimator is, and to estimate the true covariance.

Furthermore, we showed that the subsampling method also gives a biased estimate of the covariance, this time for the subsample area. The reason for this is that the $C_\ell^{\rm eff}$ estimated from a survey or simulation of finite size always lack power on the largest scales, and this lack of power is then transferred to all covariance elements through the convolution of Eq.~(\ref{Eq:avgcov-cleff}).
At least on small scales, this sounds the death knell to estimating the covariance from a single simulation larger than the survey by carefully cutting different survey areas into it. On large scales, the subsampling method actually performs quite well, especially for the auto-redshift covariance, although it performs a bit less well  for the cross-redshift covariance. The subsampling method works relatively well for the auto-redshift covariance
because the lack of power affects only the smallest multipoles and because $\Cov_\ell^\mr{SSC}$ is flatter or even turning around at low $\ell$ (see left panel of Fig. \ref{Fig:Cov_ell^SSC}), giving less weight to large scales. On the contrary for cross-redshift, $\Cov_\ell^\mr{SSC}$ is still steep at low multipoles, explaining why the subsampling lack of power affects it more. We tested setups other than those presented in Fig. \ref{Fig:subsamples-illustration}, and found that the subsampling method actually gives a very good covariance estimate in cases with a larger number of subsamples or when subsamples have a larger area.

In some cases, covariance estimation from a single simulation is unavoidable. One such case arises when  analytical covariance predictions need to be compared to a carefully crafted and CPU-expensive simulation in order to validate the prediction code. Then, the formalism developed here allows  the analytical predictions to be tweaked in order to match how biased the covariance from the simulation is. In that case, we recommend comparing the covariance of the subsample area from the prediction and the estimate from the subsampling method. Indeed, the latter performs much better than the jackknife, and the bias may even  be negligible in  large-scale cases or when there is a large enough number of subsamples.

Although this article focused on the case of cluster number counts, the formalism developed and the qualitative conclusions are also valid  for other probes, including the two-point correlation function or power spectrum for galaxy clustering and weak-lensing shear.

\subsection{Debiasing (im)possibility}

\cite{Hoffmann2015} developed a method for debiasing the standard jackknife estimator by first removing the shot-noise contribution, then debiasing the SSC contribution using the ratio of the variances of the matter density field in the survey volume and subsample volume,  $r_\sigma \equiv \sigma^2_m(V_\mr{tot})/\sigma^2_m(V_\mr{JKcell})$, and finally re-adding the shot-noise. This convoluted method is possible because they worked with 3D volumes at a single fixed redshift, but it becomes theoretically impossible in the case considered in this article of a lightcone observable, due to the redshift evolution within the bin and because we consider different bins. An effective rescaling, e.g. using $r_\sigma$ at the centre of the redshift bin, is possible and may perform in a satisfactory manner for the auto-redshift covariance; however, we see that the ratios plotted in Figs. \ref{Fig:ratio-cov-eff/true-subsampling} and \ref{Fig:ratio-cov-eff/true-jackknife} are not constant with mass and redshift. Hence, it is not possible to debias simultaneously all covariance elements together, in particular not the auto-$z$ and cross-$z$ elements at the same time.\\
We emphasize that the covariances were predicted from analytical equations, not estimated from simulations, so that the ratios plotted are exact (up to machine precision), and their variations are thus genuine.

Another debiasing method would be additive instead of multiplicative. It can be  computed as
\ba\label{Eq:bias-cov}
\nonumber \Delta \, \Cov(N_\alpha,N_\beta) &\equiv \frac{1}{\left(f^\mr{sub}_\mr{sky}\right)^2} \sum_\ell \frac{2\ell+1}{4\pi} \ \left(C_\ell(M) - C_\ell^\mr{eff}\right) \\
& \qquad \times\Cov_\ell^\mr{SSC}\left(N_\alpha , N_\beta\right)
\ea
and used  to debias the estimated covariance:
\be
\widetilde{\Cov}(N_\alpha,N_\beta) \equiv \widehat{\Cov}(N_\alpha,N_\beta) + \Delta \, \Cov(N_\alpha,N_\beta) \, .
\ee
However, the advantages of this debiasing are few: computing Eq. (\ref{Eq:bias-cov}) has exactly the same analytical requirements as computing the full prediction Eq. (\ref{Eq:Cov-counts-wcovell}), including a theoretical prediction of the halo mass function and bias (while the ratio $r_\sigma$ from \citealt{Hoffmann2015} only depends on the linear matter power spectrum and the volumes considered). As such, it is far preferable to predict the covariance entirely analytically, avoiding simulation noise and thus the need for a Hartlap's correction \citep{Hartlap2007} or a Bayesian marginalisation \citep{Sellentin2016}.

\subsection{Catalogue-based covariance estimators}

Another class of often-used covariance estimators are jackknife or bootstrap estimators based on object catalogues  instead of subsamples. These estimators are used for example for the covariance of galaxy clustering or weak-lensing correlation functions or power spectra. At first sight, a catalogue-based estimator may seem like the asymptote of the corresponding subsample-based estimator in the limit of infinitely small subsamples, but the presence of empty subsamples means that this is not the case. It also does not correspond  to the case of subsamples carefully crafted to contain a unique object;  this construction is doomed by non-uniqueness and enters the realm of a posteriori statistics.

It can be seen that catalogue-based estimators have intrinsic limitations. We  argue that these limitations  limit  their use for estimating covariance of large-scale structure observables. First, they are a tautological nonsense for number counts, e.g. a jackknife would always have $N_\mr{obj}-1$ objects and thus the estimated covariance is always zero. Though this may sound trivial, it  means that it does not permit their use for galaxy and/or cluster number counts, and thus for combining probes. Combined probes is a main objective for current and future galaxy surveys, and thus a careful prediction of the joint probe covariance, including the cross-covariance between probes, is of prime importance. Second, catalogue-based estimators give weight to dense regions, but regions devoid of galaxies will not change between jackknife samples (nor between bootstrap samples); this is problematic for cosmological applications as void properties do change with cosmological parameters. Finally, in a field where all galaxies would be grouped in a single pixel, a catalogue-based jackknife/bootstrap estimator would find the same correlation function or power spectrum in all jackknife/bootstrap samples and thus a zero covariance, which would not be the correct answer at all.

Though it is not the purpose of this article to examine thoroughly the case of catalogue-based estimators, the arguments presented above allow us to cast doubt on the validity of these estimators already for a Gaussian matter density field, and even more so for super-sample covariance.

%%%%%%%%%%%%%%%%%%%%%%%%%%%%%%%%%%%%%%%%%%%%%%%%%%%%%%%%%%%%%%%%%%%
\section{Conclusion}\label{Sect:conclusion}
We have developed the formalism necessary to predict the effect of SSC on standard covariance estimators based on spatial subsamples of the data. We demonstrated that internal covariance estimation with the jackknife estimator is significantly biased, calling into question the applicability of the method for real galaxy surveys. Another application is the case of covariance estimation using survey-sized subsamples extracted from a large simulation. We showed that in this case  the estimated covariance is also biased, due to the non-independence of the subsamples, though this bias is less important than in the jackknife case.
The main reason for these difficulties is that the large-scale modes that underlie super-sample covariance are not accessible from within the survey, and the largest ones are also inaccessible to simulations.
Although this challenges the use of single simulations for actual covariance estimation, the formalism allows the validation of theoretical prediction codes against single simulations by tweaking the predictions to reproduce the subsampling effect.
Finally, while the formalism and the qualitative conclusions from this article were developed in the case of cluster number counts, they are also valid for other probes, such as galaxy clustering or cosmic shear.

%%%%%%%%%%%%%%%%%%%%%%%%%%%%%%%%%%%%%%%%%%%%%%%%%%%%%%%%%%%%%%%%%%%%
\section*{Acknowledgements}
\vspace{0.2cm}

We thank Camille Bonvin and Kai Hoffmann for discussions that improved this article.
We acknowledge the use of the {\tt Healpix} package by \cite{Gorski2005}.
F.L. and M.K. acknowledge support by the Swiss National Science Foundation.

%%%%%%%%%%%%%%%%%%%%%%%%%%%%%%%%%%%%%%%%%%%%%%%%%%%%%%%%%%%%%%%%%%%%
\bibliographystyle{aa}
\bibliography{bibliography}
%%%%%%%%%%%%%%%%%%%%%%%%%%%%%%%%%%%%%%%%%%%%%%%%%%%%%%%%%%%%%%%%%%%%

\appendix

\section{Bootstrap asymptotes to jackknife}\label{App:bootstrap}

\subsection{With probabilities}
In the limit $N_\mr{bs} \rightarrow \infty$, the bootstrap estimator can be written as
\be
\widehat{\Cov}(N_\alpha,N_\beta) = \sum_{b} p(b) \; N_\alpha^\mr{bs}(b) \, N_\beta^\mr{bs}(b) \ - \overline{N}_\alpha^\mr{bs} \overline{N}_\beta^\mr{bs}
,\ee
where the sum runs over all $N_\mr{sub}^{N_\mr{sub}}$ possible bootstrap samples, and $p(b) = N_\mr{sub}^{-N_\mr{sub}}$. It follows that the estimator takes the form of Eq. (\ref{Eq:Cov-wij}) with
\ba
w_{i,j}^{bs} &= \sum_{k,m=1}^{N_\mr{sub}} p(b_k=i,b_m=j) \ - 1 \\
&= \sum_{k\neq m} \frac{1}{N^2_\mr{sub}} + \delta_{i,j}\sum_{k=m} \frac{1}{N_\mr{sub}} \ - 1\\
&= \delta_{i,j} - \frac{1}{N_\mr{sub}} \, .
\ea

\subsection{With symmetries}
This proof revolves around the fact that there is basically only one natural covariance estimator based on subsamples $i=1..N_\mr{sub}$. Indeed the estimator must be symmetric under permutations of the subsample indices $i$; it thus follows that there are two constants $A$ and $B$ such that
\be
w^{bs}_{i,j} = A \, \delta_{i,j} + B \, .
\ee
Furthermore, the bootstrap weights $w_{i,j}^{bs}$ have to satisfy the constraint
\be\label{Eq:constraint-wbs}
\sum_{i,j=1}^{N_\mr{sub}} w^{bs}_{i,j} = 0 \, .
\ee
This implies that
\be
w^{bs}_{i,j} \propto \delta_{i,j} - \frac{1}{N_\mr{sub}} \, .
\ee
The proportionality constant can be found by examining the case $N_\alpha(i) = \delta_{1,i}$ for which the covariance estimator gives $\widehat{\Cov}=1$.

\end{document}